# Near Quadratic Matrix Multiplication Modulo Composites

*Preliminary version*

Vince Grolmusz [*]

**Abstract**

We show how one can use non-prime-power, composite moduli for computing representations of the product of two $n \times n$ matrices using only $n^{2+o(1)}$ multiplications.

## 1 Introduction

The matrix multiplication is a basic operation in mathematics in applications in almost every branch of mathematics itself, and also in the science and engineering in general. An important problem is finding algorithms for fast matrix multiplication. The natural algorithm for computing the product of two $n \times n$ matrices uses $n^3$ multiplications. The first, surprising algorithm for fast matrix multiplication was the recursive method of Strassen [Str69], with $n^{2.81}$ multiplications. The best known algorithm today was given by Coppersmith and Winograd [CW90], requiring only $n^{2.376}$ multiplications. Some of these methods can be applied successfully in practice for the multiplication of large matrices[Bai88].

The best lower bounds for the number of needed multiplications are between $2.5n^2$ and $3n^2$, depending on the underlying fields (see [Blä99], [Bsh89], [Shp01]). A celebrated result of Raz [Raz02] is an $\Omega(n^2 \log n)$ lower bound for the number of multiplications, if only bounded scalar multipliers can be used in the algorithm.

The main result of the present paper is an algorithm with $n^{2+o(1)}$ multiplications for computing a *representation* of the matrix product modulo non-prime power composite numbers (e.g., 6). The algorithm is a straightforward application of a method of computing the representation of the dot-product of two length-$n$ vectors with $n^{o(1)}$ multiplications.

## 2 Preliminaries

In [Gro02] we gave the definition of the *a-strong (i.e., alternative-strong) representation* of polynomials. Here we define the *alternative*, and the *0-a-strong* and the *1-a-strong* representations of polynomials. Note that the 0-a-strong representation, defined here, coincides with the a-strong representation of the paper [Gro02].

Note also, that for prime or prime-power moduli, polynomials and their representations (defined below), coincide.

---
[*]Department of Computer Science, Eötvös University, Budapest, Pázmány P. stny. 1/C, H-1117 Budapest, Hungary; E-mail: grolmusz@cs.elte.hu

**Definition 1** *Let $m$ be a composite number $m = p_1^{e_1} p_2^{e_2} \cdots p_\ell^{e_\ell}$. Let $Z_m$ denote the ring of modulo $m$ integers. Let $f$ be a polynomial of $n$ variables over $Z_m$:*

$$f(x_1, x_2, \ldots, x_n) = \sum_{I \subset \{1,2,\ldots,n\}} a_I x_I,$$

*where $a_I \in Z_m$, $x_I = \prod_{i \in I} x_i$. Then we say that*

$$g(x_1, x_2, \ldots, x_n) = \sum_{I \subset \{1,2,\ldots,n\}} b_I x_I,$$

*is an*

- *alternative representation of $f$ modulo $m$, if*

$$\forall I \subset \{1, 2, \ldots, n\} \quad \exists j \in \{1, 2, \ldots, \ell\} : \quad a_I \equiv b_I \pmod{p_j^{e_j}};$$

- *0-a-strong representation of $f$ modulo $m$, if it is an alternative representation, and, furthermore, if for some $i$, $a_I \not\equiv b_I \pmod{p_i^{e_i}}$, then $b_I \equiv 0 \pmod{p_i^{e_i}}$;*

- *1-a-strong representation of $f$ modulo $m$, if it is an alternative representation, and, furthermore, if for some $i$, $a_I \not\equiv b_I \pmod{p_i^{e_i}}$, then $a_I \equiv 0 \pmod{m}$;*

**Example 2** *Let $m = 6$, and let $f(x_1, x_2, x_3) = x_1 x_2 + x_2 x_3 + x_1 x_3$, then*

$$g(x_1, x_2, , x_3) = 3 x_1 x_2 + 4 x_2 x_3 + x_1 x_3$$

*is a 0-a-strong representation of $f$ modulo 6;*

$$g(x_1, x_2, , x_3) = x_1 x_2 + x_2 x_3 + x_1 x_3 + 3 x_1^2 + 4 x_2$$

*is a 1-a-strong representation of $f$ modulo 6;*

$$g(x_1, x_2, , x_3) = 3 x_1 x_2 + 4 x_2 x_3 + x_1 x_3 + 3 x_1^2 + 4 x_2$$

*is an alternative representation modulo 6.*

In other words, for modulus 6, in the alternative representation, each coefficient is correct either modulo 2 or modulo 3, but not necessarily both.

In the 0-a-strong representation, the 0 coefficients are always correct both modulo 2 and 3, the non-zeroes are allowed to be correct either modulo 2 or 3, and if they are not correct modulo one of them, say 2, then they should be 0 mod 2.

In the 1-a-strong representation, the non-zero coefficients of $f$ are correct for both moduli in $g$, but the zero coefficients of $f$ can be non-zero either modulo 2 or modulo 3 in $g$, but not both.

We considered elementary symmetric polynomials

$$S_n^k = \sum_{\substack{I \subset \{1,2,\ldots,n\} \\ |I|=k}} \prod_{i \in I} x_i$$

in [Gro02], and proved that for constant $k$'s, 0-a-strong representations of elementary symmetric polynomials $S_n^k$ can be computed dramatically faster over non-prime-power composites than over primes: we gave a depth-3 multilinear arithmetic circuit with sub-polynomial number of multiplications (i.e., $n^\varepsilon, \forall \varepsilon > 0$), while over fields or prime moduli computing these polynomials on depth-3 multilinear circuits needs polynomial (i.e., $n^{\Omega(1)}$) multiplications.

In particular, we proved the following theorem:

**Theorem 3 ([Gro02])** (i) Let $m = p_1 p_2$, where $p_1 \neq p_2$ are primes. Then a degree-2 0-a-strong representation of

$$S_n^2(x,y) = \sum_{\substack{i,j \in \{1,2,\ldots,n\} \\ i \neq j}} x_i y_j,$$

modulo $m$:

$$\sum_{\substack{i,j \in \{1,2,\ldots,n\} \\ i \neq j}} a_{ij} x_i y_j \qquad (1)$$

can be computed on a bilinear $\Sigma\Pi\Sigma$ circuit of size

$$\exp(O(\sqrt{\log n \log \log n})).$$

Moreover, this representation satisfies that $\forall i \neq j : a_{ij} = a_{ji}$.

(ii) Let the prime decomposition of $m = p_1^{e_1} p_2^{e_2} \cdots p_r^{e_r}$. Then a degree-2 0-a-strong representation of $S_n^2(x,y)$ modulo $m$ of the form (1) can be computed on a bilinear $\Sigma\Pi\Sigma$ circuit of size

$$\exp\left(O\left(\sqrt[r]{\log n (\log \log n)^{r-1}}\right)\right).$$

Moreover, this representation satisfies that $\forall i \neq j : a_{ij} = a_{ji}$.

□

**Corollary 4** *The 0-a-strong representation of (1) can be computed using*

$$\exp(O(\sqrt{\log n \log \log n}))$$

*multiplications.*

**Proof:** The proof immediately follows from Theorem 3, and the definition of $\Sigma\Pi\Sigma$ circuits, given in [Gro02]. □

Now we prove the following

**Theorem 5** (i) Let $m = p_1 p_2$, where $p_1 \neq p_2$ are primes. Then a degree-2 1-a-strong representation of the dot-product

$$f(x_1, x_2, \ldots, x_n, y_1, y_2, \ldots, y_n) = \sum_{i=1}^n x_i y_i$$

can be computed with

$$\exp(O(\sqrt{\log n \log \log n})) \qquad (2)$$

multiplications.

(ii) Let the prime decomposition of $m = p_1^{e_1} p_2^{e_2} \cdots p_r^{e_r}$. Then a degree-2 1-a-strong representation of the dot-product $f$ modulo $m$ can be computed using

$$\exp\left(O\left(\sqrt[r]{\log n (\log \log n)^{r-1}}\right)\right) \qquad (3)$$

multiplications.

(iii) Moreover, the representations of (i) and (ii) can be computed on bilinear $\Sigma\Pi\Sigma$ circuits of size (2), and (3), respectively.

**Proof:** Let $g(x,y) = g(x_1, x_2, \ldots, x_n, y_1, y_2, \ldots, y_n)$ be the degree-2 polynomial from Theorem 3 which is a 0-a-strong representation of $S_n^2(x,y)$. Then consider polynomial

$$h(x,y) = (x_1 + x_2 + \ldots + x_n)(y_1 + y_2 + \ldots + y_n) - g(x,y).$$

In $h(x,y)$, the coefficients of monomials $x_i y_i$ are all 1's modulo $m$, and the coefficients of monomials $x_i y_j$, for $i \neq j$ are 0 at least for one prime-power divisor of $m$. Consequently, by Definition 1, $h(x,y)$ is a 1-a-strong representation of the dot-product $f(x,y)$. □

In contrast, as we proved in [Gro02], the 0-a-strong representation of the dot-product cannot be computed with few multiplications:

**Theorem 6 ([Gro02])** *Let*

$$f(x_1, x_2, \ldots, x_n, y_1, y_2, \ldots, y_n) = \sum_{i=1}^{n} x_i y_i$$

*be the inner product function. Suppose that a $\Sigma\Pi\Sigma$ circuit computes an a-strong representation of $f$ modulo 6. Then the circuit must have at least $\Omega(n)$ multiplication gates.*

□

**Definition 7** *Let $A = \{a_{ij}\}$ and $B = \{b_{ij}\}$ be two $n \times n$ matrices over $Z_m$. Then $C = \{c_{ij}\}$ is the alternative (1-a-strong, 0-a-strong) representation of the product-matrix $AB$, if for $1 \leq i, j \leq n$, $c_{ij}$ is an alternative (1-a-strong, 0-a-strong) representation of polynomial*

$$\sum_{k=1}^{n} a_{ik} b_{kj}$$

*modulo $m$, respectively.*

Our main theorem here is

**Theorem 8** (i) *Suppose that $m$ has two distinct prime divisors. Then a 1-a-strong representation modulo $m$ of $AB$ can be computed using*

$$n^2 2^{O(\sqrt{\log n (\log \log n)})} = n^{2+o(1)} \tag{4}$$

*multiplications.*

(ii) *Suppose that $m$ has $r$ distinct prime divisors. Then a 1-a-strong representation modulo $m$ of $AB$ can be computed using*

$$n^2 2^{O(\sqrt[r]{\log n (\log \log n)^{r-1}})} = n^{2+o(1)} \tag{5}$$

*multiplications.*

(iii) *Moreover, the representations of (i) and (ii) can be computed on bilinear $\Sigma\Pi\Sigma$ circuits of size (4), and (5), respectively.*

The proof is immediate by applying $n^2$-times the representation of the dot-product, implied by Theorem 5.□

## 3  On 1-a-strong representations

The following lemma describes some useful properties of the 1-a-strong representations. Before stating it, we need a

**Definition 9** *Let $g$ be the 1-a-strong representation of polynomial $f$ modulo $m = p_1^{e_1} p_2^{e_2} \cdots p_\ell^{e_\ell}$. Then those monomials, which appears in $g$, but not in $f$, with a non-zero coefficient modulo $m$ are called* surplus *monomials. Let $g'$ be the 1-a-strong representation of polynomial $f'$ modulo $m$. We say, that $g$ and $g'$ have disjoint surpluses, if the set of the monomials in their surpluses are disjoint. We say that they have compatible surpluses, if a monomial $x_I$ is present with coefficient $a_I$ in the surplus of $g$ and with coefficient $a'_I$ in the surplus of $g'$, then there exist an $i : 1 \leq i \leq \ell$, such that*

$$a_I \equiv 0 \equiv a'_I \pmod{p_i^{e_i}}.$$

Note, that the coefficients of the surplus monomials are zero modulo at least one prime-power divisor of $m$. The reason of defining disjoint and compatible surplus is the fact, that the sum of surplus monomials may have a non-zero coefficient for all prime divisors of $m$, e.g., if $3xy$ can be a surplus mod 6, $4xy$ can be a surplus mod 6, but their sum, $xy$ cannot be a surplus modulo 6. Consequently, in general, the sum of the 1-a-strong representations is not a 1-a-strong representation of the sum of the original polynomials.

**Lemma 10** *Let $f$ and $f'$ be polynomials over the ring $Z_m$, and let $g$ and $g'$ their 1-a-strong representations modulo $m$. Then*

 *(i) $gg'$ is a 1-a-strong representation of $ff'$ modulo $m$.*

 *(ii) Suppose, that $g$ and $g'$ have either disjoint or compatible surpluses modulo $m$. Then $g + g'$ is a 1-a-strong representation modulo $m$.*

**Proof:**    If $a_I x_I$ is a surplus monomial, then its product with anything else will be zero $\pmod{p_i^{e_i}}$ for some $i$, since $a_I \equiv 0 \pmod{p_i^{e_i}}$. If $x_I$ is not a surplus in $g$ and $x_J$ is not a surplus in $g'$, then their coefficients are equal to the corresponding coefficients in $f$ and $f'$, respectively, thus their product may have a zero coefficient in $gg'$, but then the corresponding coefficient is also zero in $ff'$; this proves (i). If $a_I$ is zero $\pmod{p_i^{e_i}}$, and $a'_I$ is zero $\pmod{p_i^{e_i}}$, then $a_I + a'_I$ is also zero $\pmod{p_i^{e_i}}$, this proves (ii). □

## 4  Open problems

It would be interesting to compute 0-a-strong representations of the matrix product or the matrix-vector product using fewer multiplications than the currently best known algorithms for computing the exact values.

**Acknowledgment.**
   The author acknowledges the partial support of an NKFP grant and an ETIK grant and the EU FP 5 grant IST FET IST-2001-32012.

*Grolmusz: Near Quadratic Matrix Multiplication Modulo Composites* 6*Grolmusz: Near Quadratic Matrix Multiplication Modulo Composites* 6

# References

[Bai88] David H. Bailey. Extra high speed matrix multiplication on the Cray-2. *SIAM J. Sci. Statist. Comput.*, 9(3):603–607, 1988.

[Blä99] Markus Bläser. A $\frac{5}{2}n^2$-lower bound for the rank of $n \times n$-matrix multiplication over arbitrary fields. In *40th Annual Symposium on Foundations of Computer Science (New York, 1999)*, pages 45–50. IEEE Computer Soc., Los Alamitos, CA, 1999.

[Bsh89] Nader H. Bshouty. A lower bound for matrix multiplication. *SIAM J. Comput.*, 18(4):759–765, 1989.

[CW90] Don Coppersmith and Shmuel Winograd. Matrix multiplication via arithmetic progressions. *J. Symbolic Comput.*, 9(3):251–280, 1990.

[Gro02] Vince Grolmusz. Computing elementary symmetric polynomials with a sub-polynomial number of multiplications. Technical Report TR02-052, ECCC, 2002. ftp://ftp.eccc.uni-trier.de/pub/eccc/reports/2002/TR02-052/index.html.

[Raz02] Ran Raz. On the complexity of matrix product. In *Proceedings of the thirty-fourth annual ACM symposium on Theory of computing*. ACM Press, 2002.

[Shp01] Amir Shpilka. Lower bounds for matrix product. In *IEEE Symposium on Foundations of Computer Science*, pages 358–367, 2001.

[Str69] V. Strassen. Gaussian elimination is not optimal. *Numerische Mathematik*, 13:354–356, 1969.